# Valence bond glass state in the 4$d^1$ fcc antiferromagnet Ba$_2$LuMoO$_6$

O. H. J. Mustonen[1], H. M. Mutch[1], H. C. Walker[2], P. J. Baker[2], F. C. Coomer[3], R. S. Perry[4], C. Pughe[1], G. B. G. Stenning[2], C. Liu[5], S. E. Dutton[5] and E. J. Cussen[1*]

[1]*Department of Material Science and Engineering, University of Sheffield, Mappin Street, Sheffield, S1 3JD, United Kingdom*

[2]*ISIS Pulsed Neutron and Muon Source, STFC Rutherford Appleton Laboratory, Harwell Campus, Didcot, OX11 0QX, United Kingdom*

[3]*Johnson Matthey Battery Materials, Blount's Court, Sonning Common, Reading, RG4 9NH United Kingdom*

[4]*London Centre for Nanotechnology and Department of Physics and Astronomy, University College London, Gower Street, London, WC1E 6BT, United Kingdom*

[5]*Cavendish Laboratory, University of Cambridge, Cambridge CB3 0HE, United Kingdom*

[*] Corresponding author
e.j.cussen@sheffield.ac.uk

**Abstract**

*B*-site ordered 4$d^1$ and 5$d^1$ double perovskites have a number of potential novel ground states including multipolar order, quantum spin liquids and valence bond glass states. These arise from the complex interactions of spin-orbital entangled $J_{eff}$ = 3/2 pseudospins on the geometrically frustrated fcc lattice. The 4$d^1$ Mo$^{5+}$ perovskite Ba$_2$YMoO$_6$ has been suggested to have a valence bond glass ground state. Here we report on the low temperature properties of powder samples of isostructural Ba$_2$LuMoO$_6$: the only other known cubic 4$d^1$ perovskite with one magnetic cation. Our muon spectroscopy experiments show that magnetism in this material remains dynamic down to 60 mK without any spin freezing or magnetic order. A singlet-triplet excitation with a gap of Δ = 28 meV is observed in inelastic neutron scattering. These results are interpreted as a disordered valence bond glass ground state similar to Ba$_2$YMoO$_6$. Our results highlight the differences of the 4$d^1$ double perovskites in comparison to cubic 5$d^1$ analogues, which have both magnetic and multipolar order.

## INTRODUCTION

Recently, 4d and 5d transition metal compounds have been widely investigated as potential hosts for novel quantum states of matter[1–4]. The origin of the rich physics of these heavy transition metal compounds lies in the interplay of strong electron correlation effects (Hubbard $U$) and significant spin-orbit coupling $\lambda$[1]. This can result in entanglement of the spin and orbital degrees of freedom. An important example of this is $Sr_2IrO_4$, in which the $5d^5$ $Ir^{4+}$ cations have a spin-orbital entangled $J_{eff} = 1/2$ state leading to insulating behavior[5]. Moreover, honeycomb lattice iridates and α-$RuCl_3$ with $J_{eff} = 1/2$ pseudospins have emerged as possible realizations of Kitaev's exactly solvable quantum spin liquid model[3,6]. While these $4d^5$ and $5d^5$ $J_{eff} = 1/2$ systems have garnered the most attention, novel physics are also observed in other 4d and 5d materials[2]. The $5d^5$ and $4d^5$ low-spin systems have one hole on the $t_{2g}$ orbitals. A mirror of this situation occurs in $4d^1$ and $5d^1$ compounds with one electron on the $t_{2g}$ orbitals.

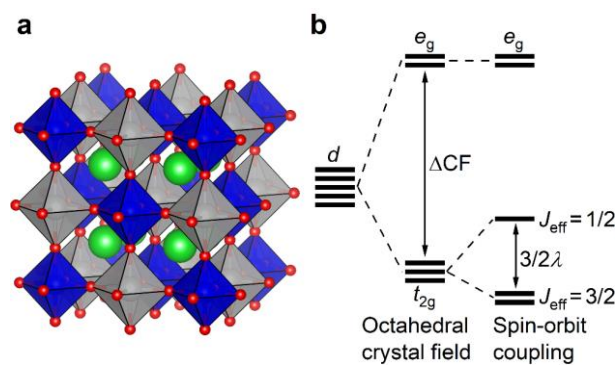

**Fig. 1. Structure and orbital splitting**. **a** The double perovskite structure of $Ba_2LuMoO_6$. Ba, Lu, Mo and O are represented by the green, gray, blue and red spheres, respectively. The magnetic $Mo^{5+}$ ($4d^1$) cations in blue form an undistorted fcc lattice. **b** Scheme of the orbital splitting for a $4d^1$ or a $5d^1$ cation adapted from refs. [5,7]. The octahedral crystal field splits the d orbitals into $t_{2g}$ and $e_g$ states. The $t_{2g}$ states have an effective orbital angular momentum $L_{eff} = 1$. Spin-orbit coupling further splits the six-fold degenerate $t_{2g}$ states into a $J_{eff} = 3/2$ quartet ground state and a $J_{eff} = 1/2$ doublet excited state. Note that for $4d^5$ and $5d^5$ cations the situation is reversed and $J_{eff} = 1/2$ is the ground state while $J_{eff} = 3/2$ is the excited state.

Cubic $A_2B'B''O_6$ double perovskites, where the only magnetic cation is a $4d^1$ or a $5d^1$ transition metal on the $B''$ site, have a variety of potential unusual ground states[1,2]. The structure, as shown in Fig. 1a, consists of corner-sharing octahedra, where the B-site cations alternate in a rocksalt-type order forming an fcc lattice of the $B''$ $d^1$ cations[8]. The octahedral crystal field splits the d-orbitals into six $t_{2g}$ states and four $e_g$ states. When the crystal field splitting is large enough to prevent $t_{2g}$-$e_g$ mixing, the $t_{2g}$ states can be described as having effective orbital angular momentum of $L_{eff} = 1$. Spin-orbit coupling, as observed in 4*d* and 5*d* transition metal compounds, further splits the $t_{2g}$ orbitals into spin-orbital entangled *J* = *L*+*S* states: a $J_{eff}$ = 3/2 quartet ground state and a $J_{eff}$ = 1/2 doublet excited state (Fig. 1b)[7]. The $J_{eff}$ = 3/2 ground state is nominally nonmagnetic with *M* = 2*S* - *L* = 0 as the spin and orbital moments oppose and cancel out. In real compounds, the orbital moment is reduced by hybridization with oxygen leading to a small overall moment[2]. Conversely, in the excited $J_{eff}$ = 1/2 state the spin and orbital moments add up to a larger moment.

The $J_{eff}$ = 3/2 pseudospins and their complex interactions on the geometrically frustrated fcc lattice of $d^1$ double perovskites give rise to rich physics in these materials. This can enable bond-directional exchange on the fcc lattice similar to Kitaev interactions on the honeycomb lattice[9–11]. Theoretical models of the $d^1$ fcc systems predict novel ground states including spin liquid states[9,10,12], multipolar order[12] and valence bond glass states[13]. In terms of materials, the $5d^1$ double perovskites $Ba_2NaOsO_6$[14–20] and $Ba_2MgReO_6$[21–24] have been widely investigated for possible multipolar order. While both compounds have magnetically ordered ground states, there is indirect evidence of quadrupolar ordering above the magnetic ordering transition.

The most studied $4d^1$ double perovskite is $Ba_2YMoO_6$[25–30]. It has a fcc lattice of $4d^1$ $Mo^{5+}$ cations with a $J_{eff}$ = 3/2 ground state. No structural distortions from cubic $Fm\bar{3}m$ symmetry are observed down to 3 K[31], although a Jahn-Teller distortion would be expected in $d^1$ compounds even in the presence of strong spin-orbit coupling[32]. Despite strong antiferromagnetic interactions with $\Theta_{CW}$ = -160 K, muon spin rotation and relaxation measurements show that $Ba_2YMoO_6$ does not magnetically order, although a partial spin glass transition occurs at 1.3 K[29]. NMR and specific heat measurements suggested the presence of spin singlets in the ground state[26,31]. Inelastic neutron scattering measurements revealed a gapped singlet-triplet excitation with Δ = 28 meV[28].

These experimental results for $Ba_2YMoO_6$ have been interpreted as a valence bond glass ground state, where nonmagnetic spin singlets gradually form in a disordered fashion as the temperature is decreased, while some orphan spins remain paramagnetic[13,26,29]. $Ba_2YMoO_6$ has also been suggested to be a spin liquid candidate[9,10]. The proposed spin liquid has specific power law scaling in susceptibility and heat capacity, which could be used to distinguish it from a valence bond glass state[10]. Both interpretations can explain the observed inelastic neutron scattering data[10,13], while the valence bond glass better explains the glassy behavior observed in AC susceptibility around 50 K[13,26]. However, the partial spin glass transition at 1.3 K in $Ba_2YMoO_6$ is not expected for the valence bond glass nor the spin liquid. This raises the question whether a better $d^1$ candidate material for either of these novel ground states could be found.

$Ba_2LuMoO_6$ is the only other known cubic $4d^1$ double perovskite with one magnetic cation, but its ground state is not known[33]. Like $Ba_2YMoO_6$, it retains the $Fm\bar{3}m$ symmetry with an fcc lattice of

$4d^1$ Mo$^{5+}$ cations down to 2 K[33]. In this article, we report on the low-temperature properties and possible ground states of Ba$_2$LuMoO$_6$. Our muon spin rotation and relaxation measurements reveal a lack of magnetic order or spin freezing down to 60 mK. Inelastic neutron scattering measurements show a gapped magnetic excitation with Δ = 28 meV, which is interpreted as a singlet-triplet excitation. The presence of both spin singlets and dynamic magnetism is interpreted as a valence bond glass state similar to Ba$_2$YMoO$_6$, but without freezing of orphan spins. Our work highlights the differences between the cubic $4d^1$ double perovskites Ba$_2$LuMoO$_6$ and Ba$_2$YMoO$_6$ and their $5d^1$ analogues Ba$_2$NaOsO$_6$ and Ba$_2$MgReO$_6$, which have both magnetic and quadrupolar order.

**RESULTS**

**X-ray diffraction**

Phase purity and crystal structure of Ba$_2$LuMoO$_6$ samples were investigated using X-ray powder diffraction. The sample was phase pure without any impurity peaks in the diffraction pattern (Supplementary Fig. 1). The structure of Ba$_2$LuMoO$_6$ was found to be $Fm\bar{3}m$ in excellent agreement with previous neutron diffraction results[33] (Supplementary Table 1). The most common type of structural disorder observed in double perovskites is antisite disorder of the *B*-sites[8]. Our refined site occupancies are 0.99(1) Mo and 0.01(1) Lu on the Mo-site with 0.99(1) Lu and 0.01(1) Mo on the Lu-site. This shows Ba$_2$LuMoO$_6$ is a highly ordered double perovskite without significant antisite disorder similar to other Mo$^{5+}$ double perovskites[25,27].

**Magnetic susceptibility**

The magnetic properties of Ba$_2$LuMoO$_6$ were investigated using an MPMS3 SQUID magnetometer. Magnetic susceptibility of Ba$_2$LuMoO$_6$ (Fig. 2a) exhibits typical paramagnetic behaviour. The zero-field cooled (ZFC) and field cooled (FC) curves overlap, therefore only the former is shown. The inverse magnetic susceptibility reveals two regions: a high-temperature Curie-Weiss region as expected and an additional low-temperature linear region. The change in slope between these two regions occurs at ≈50 K. Curie-Weiss fits to the high-temperature region between 200-300 K yielded a Curie-Weiss constant of $\Theta_{CW}$ = -114(1) K, indicating significant antiferromagnetic interactions between Mo$^{5+}$ cations. The effective paramagnetic moment of Mo$^{5+}$ was found to be $\mu_{eff}$ = 1.32(1) $\mu_B$, which is lower than the 1.73 $\mu_B$ expected for a *S* = 1/2 cation. This reduced moment could be evidence of an unquenched orbital moment as expected in the $J_{eff}$ = 3/2 state or strong quantum fluctuations[12,14]. The low-temperature inverse susceptibility was fitted in the range 2-20 K yielding $\Theta_{CW}$ = -1.9(2) K and $\mu_{eff}$ = 0.67(1) $\mu_B$.

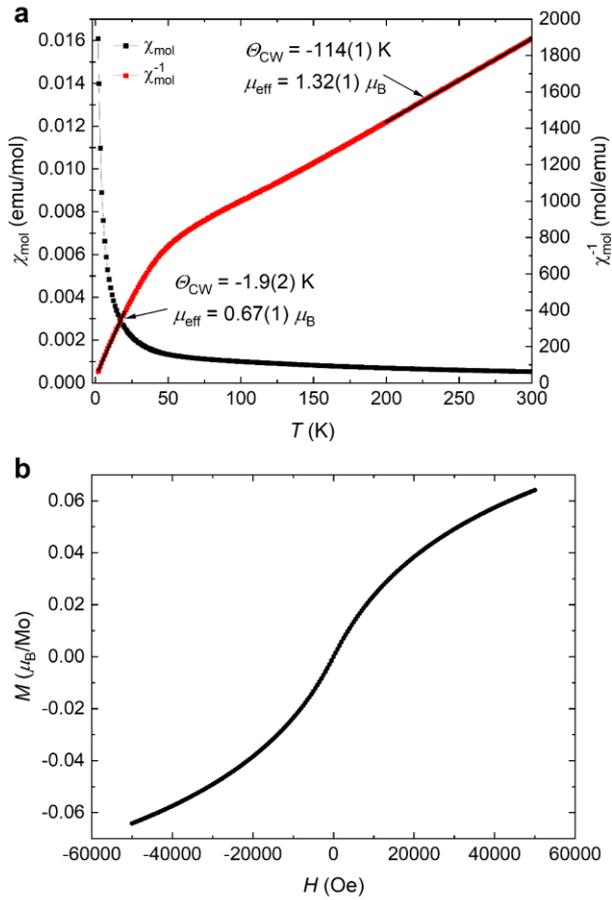

**Fig. 2. Magnetic susceptibility**. **a** DC magnetic susceptibility (black) and inverse molar magnetic susceptibility (red) of $Ba_2LuMoO_6$ as a function of temperature. Only the ZFC data is shown as the ZFC and FC curves overlap. The magnetic susceptibility appears paramagnetic with no transitions observed down to 2 K. In the inverse susceptibility two linear temperature regimes are observed: one at high temperatures (as expected) and one at low temperatures. **b** Field-dependent magnetization curve at 2 K. The $M(H)$ curve has an S shape without any hysteresis as expected for paramagnetic materials at low temperatures.

The magnetic behaviour of $Ba_2LuMoO_6$ is very similar to $Ba_2YMoO_6$., which also has two linear Curie-Weiss regions in the inverse susceptibility. The overall antiferromagnetic interactions in $Ba_2LuMoO_6$ are somewhat weaker than in $Ba_2YMoO_6$ with Curie-Weiss constants $\Theta_{CW}$ = -114 K and $\Theta_{CW}$ = -160 K, respectively[26]. The high-temperature effective paramagnetic moments are around 1.4 $\mu_B$ in both compounds. The low-temperature susceptibility is similar in both $Ba_2LuMoO_6$ and $Ba_2YMoO_6$, and the change in slope occurs around 50 K for both compounds. This feature in the inverse susceptibility of $Ba_2YMoO_6$ and related $Ba_{2-x}Sr_xYMoO_6$ phases has been interpreted as the gradual formation of valence bond singlets[26,27]. It should be noted that the change in inverse susceptibility occurs near 50 K[26], but spin singlets in $Ba_2YMoO_6$ were observed up to 125 K in inelastic neutron scattering[28].

Another possible explanation for the magnetic susceptibility is that at low temperatures the $Mo^{5+}$ cations are in the $J_{eff}$ = 3/2 ground state with a low moment, but at higher temperatures some electrons are excited to the high-moment $J_{eff}$ = 1/2 state. This is supported by the low-temperature effective moments of ≈0.6-0.7 $\mu_B$, which are very similar to those of $5d^1$ double perovskites known to have a $J_{eff}$ = 3/2 ground state. On the other hand, this excitation gap is far too large for thermal excitations at relevant temperatures as the spin-orbit coupling constant $\lambda$ = 128 meV for $Mo^{5+}$ corresponds to 1500 K[31,34]. Recently, it has been proposed[35] that a dynamical Jahn-Teller effect could mix the $J_{eff}$ = 3/2 and $J_{eff}$ = 1/2 states explaining the observed change in the effective paramagnetic moment. Nevertheless, it is clear that magnetic susceptibility alone is insufficient evidence for a valence bond glass state.

The field-dependent magnetization $M(H)$ curve at 2 K is shown in Fig. 2b. The curve has an S shape without any hysteresis and relatively low magnetization of less than 0.1 $\mu_B$/Mo at 50 kOe. The field-dependent data is in qualitative agreement with the Brillouin function behavior expected of paramagnets at low temperatures. The low-temperature $M(H)$ data are similar to those reported for $Ba_2YMoO_6$[26].

**Specific heat**

The specific heat of $Ba_2LuMoO_6$ was measured using a thermal relaxation method in order to investigate possible phase transitions such as magnetic ordering (Fig. 3). A small increase in $C_p/T$ is observed at low temperatures, similar to $Ba_2YMoO_6$[26]. The specific heat data for $Ba_2LuMoO_6$ does not contain any sharp lambda anomalies expected for magnetic ordering transitions. Despite the small magnetic moment of the $4d^1$ and $5d^1$ systems, lambda anomalies have been observed in magnetically ordered double perovskites such as $La_2LiMoO_6$ and $Ba_2MgReO_6$[22,36]. $Ba_2MgReO_6$ also has a quadrupolar ordering transition $T_q$ above $T_N$, which can be detected in the specific heat as a broader peak[22]. We do not observe any such peak in the specific heat data for $Ba_2LuMoO_6$. This suggests that there are no magnetic transitions in $Ba_2LuMoO_6$ at least down to 2 K.

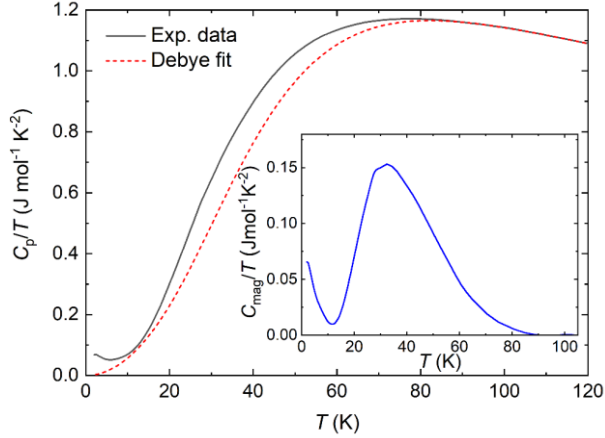

**Fig. 3. Specific heat.** Zero-field specific heat of $Ba_2LuMoO_6$ as function of temperature (black line) and the estimated lattice specific heat from Debye fits above 90 K (red dash line). No lambda anomalies indicative of magnetic order are observed down to 2 K. Moreover, there is no broad peak in specific heat associated with possible quadrupolar order in related materials. Inset: Estimated magnetic specific heat of $Ba_2LuMoO_6$. A broad maximum is observed around 30 K.

The specific heat of a magnetic compound consists of its magnetic specific heat and the lattice contribution to specific heat. The lattice contribution was estimated by fitting two Debye functions to the high-temperature data above 90 K with Debye temperatures 275 K and 956 K. The same approach has been previously used for $Ba_2MgReO_6$[22]. The estimated magnetic specific heat of $Ba_2LuMoO_6$ is shown in the Fig. 3 inset. The magnetic specific heat has a broad maximum around 30 K. No clear magnetic transition is observed. Integrating up to 90 K results in a magnetic entropy of 5.5 J K$^{-1}$ mol$^{-1}$. This is 48% of the expected entropy for a $J_{eff}$ = 3/2 system, but close to that of a $S$ = 1/2 system. It is common for magnetic materials that do not order to retain significant spin entropy at low temperatures[37]. Moreover, the magnetic specific heat and integrated entropy both have significant uncertainties due to the difficulty of estimating and subtracting the lattice contribution. For these reasons, we cannot confidently determine the spin state of the $4d^1$ Mo electron based on the specific heat.

**Muon spin rotation and relaxation**

Muon spin rotation and relaxation (μSR) experiments were performed to investigate the magnetic ground state of $Ba_2LuMoO_6$. Muons are a highly sensitive local probe of both static and dynamic magnetism[38]. In a magnetically ordered material below $T_N$, spontaneous oscillations in the measured positron asymmetry in zero-field (ZF) μSR arise from the precession of the muon spin in the static local field. These are observed in the magnetically ordered related Mo$^{5+}$ double perovskite

La$_2$LiMoO$_6$[31]. In a spin glass, the muon spins feel a distribution of static local fields below the freezing transition. For any type of static magnetism, the asymmetry should return to 1/3$^{rd}$ of the initial asymmetry at high counting times[38].

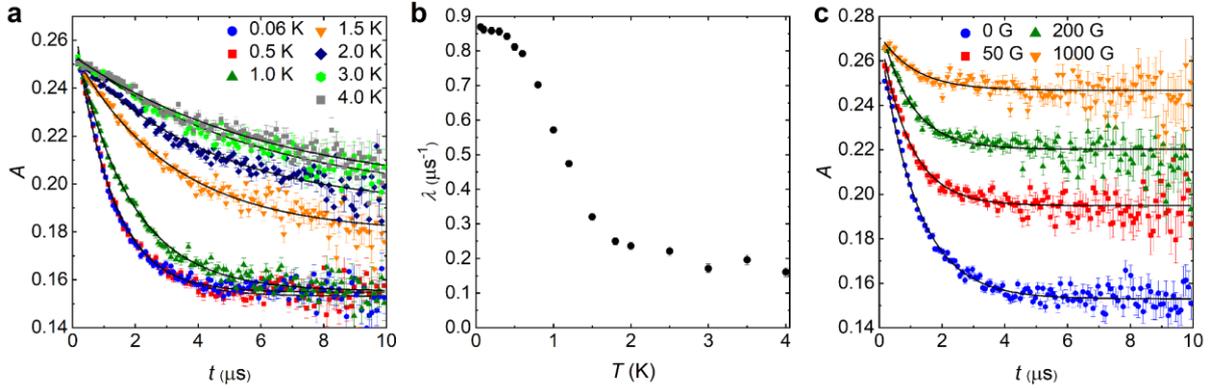

Fig. 4. Muon spin rotation and relaxation (µSR). a Zero-field µSR of Ba$_2$LuMoO$_6$ at different temperatures. Magnetism in Ba$_2$LuMoO$_6$ remains dynamic down to 60 mK as no magnetic transitions or spin freezing are observed. Muon spin relaxation of the sample can be described with a simple exponential $Ae^{-\lambda t}$. b Muon spin rotation and relaxation rate $\lambda$ as a function of temperature. Upon cooling below 1.5 K, $\lambda$ starts to rapidly increase. This shows that the internal dynamic magnetic fields slow down. Below 300 mK the relaxation rate stays constant, which indicates that the ground state is dynamic. c Longitudinal-field muon spin rotation and relaxation of Ba$_2$LuMoO$_6$ at 100 mK. Decoupling the muon spins from the internal magnetic fields required a field of over 1000 G, confirming that they are related to electronic and not nuclear spins.

We do not observe any signatures of magnetic ordering or spin freezing in Ba$_2$LuMoO$_6$ in the ZF-µSR data down to 60 mK (Fig. 4a). Instead, simple exponential relaxation behavior, typical of dynamic magnetic systems is observed. The time resolution at pulsed muon sources such as ISIS is not always sufficient to detect oscillations in magnetically ordered materials. However, if Ba$_2$LuMoO$_6$ was magnetically ordered, the static local fields would still strongly depolarize the muon spins leading to a drop in the initial asymmetry below $T_N$[38]. This is not observed. We also do not observe the 1/3$^{rd}$ tail typical of ordered materials or spin glasses. The lack of any magnetic transitions or spin freezing is consistent with both the proposed valence bond glass and spin liquid states. In a valence bond glass, the dynamic magnetism would be related to the paramagnetic orphan spins. The main component of the VBG state, the disordered spin singlets, cannot be observed by µSR as they are nonmagnetic.

The zero-field µSR between 60 mK and 4 K could be fitted using a simple exponential relaxation function:

$$A(t) = A_{exp}e^{-\lambda t} + A_{flat} \qquad (1)$$

where $A$ is the total asymmetry, $A_{exp}$ is the asymmetry of the relaxing component, $\lambda$ is the muon spin relaxation rate and $A_{flat}$ is a flat background term from the silver sample holder. This flat background term was unusually high due to sample settling in the sample holder during measurement. The fitted muon spin relaxation rate $\lambda$ is plotted as a function of temperature in Fig. 4b. The relaxation rate is inversely proportional to the fluctuation frequency of the dynamic magnetic fields. The relaxation rate starts to increase as the temperature is lowered below 2 K. This shows that the dynamic magnetic fields in $Ba_2LuMoO_6$ slow down as the temperature decreases as expected. Below 300 mK, the relaxation rate plateaus and stays constant down to at least 60 mK. This plateau in relaxation rate is a common feature of quantum spin liquid candidates such as $Zn_xCu_{4-x}(OH)_6Cl_2$[39], $Zn_xCu_{4-x}(OH)_6Cl_2$[40] and $Sr_2CuTe_{0.5}W_{0.5}O_6$.[41,42]. If $Ba_2LuMoO_6$ had a spin glass or a magnetic transition, the relaxation rate would peak at the transition temperature and no plateau would be observed. The plateau in muon spin relaxation, in addition to the lack of 1/3$^{rd}$ tail or drop in initial asymmetry down to 60 mK, confirms that $Ba_2LuMoO_6$ has a dynamic magnetic ground state ruling out magnetic order or spin freezing.

Longitudinal field (LF) µSR was measured at 100 mK in fields up to 2000 G (Fig. 4c). In LF-µSR, as the applied longitudinal field is increased, the muon spins start to decouple from the internal magnetic fields. This decoupling is observed as an increasing flat background, and the asymmetry becomes entirely flat when the muon spins are fully decoupled from the internal fields. Weak nuclear fields can be completely decoupled with low fields of 20-50 G. In $Ba_2LuMoO_6$, fully decoupling the muon spins required fields larger than 1000 G. This shows that the dynamic magnetism observed in the ZF-µSR measurements is of electronic origin and not due to nuclear magnetic moments.

The muon spin rotation and relaxation measurements reveal significant differences between $Ba_2LuMoO_6$ and $Ba_2YMoO_6$. For $Ba_2LuMoO_6$, the ZF-µSR data can be described with exponential relaxation, similar to many quantum spin liquid candidates. This dynamic magnetism is also consistent with the paramagnetic orphan spins expected in a valence bond glass state. In the case of $Ba_2YMoO_6$, two separate muon environments were proposed: a non-magnetic one with weak exponential relaxation and a magnetic one described with a stretched exponential[29]. The nonmagnetic muon environment in $Ba_2YMoO_6$ is related to the spin singlets, while the magnetic environment is related to paramagnetic orphan spins. A transition in the magnetic environment was observed at 1.3 K, which was interpreted as freezing of the orphan spins in $Ba_2YMoO_6$. The absence of this spin freezing makes $Ba_2LuMoO_6$ a more promising valence bond glass candidate than $Ba_2YMoO_6$.

## Inelastic neutron scattering

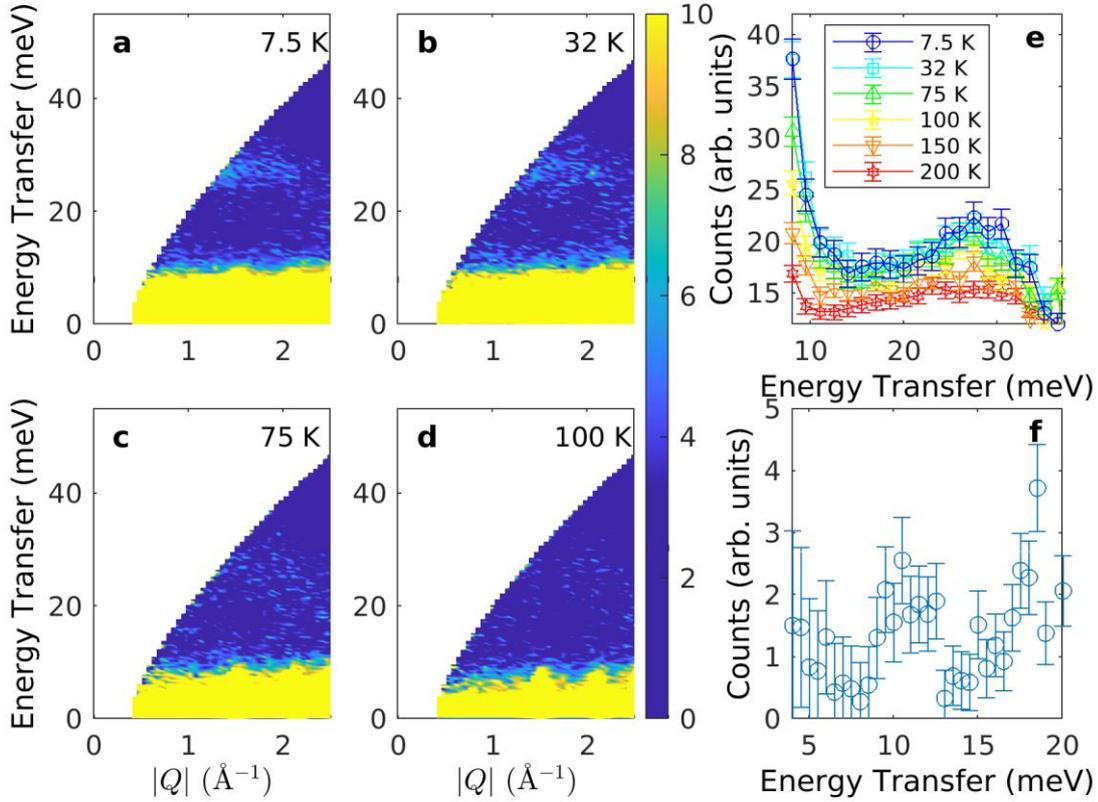

**Fig. 5. Inelastic neutron scattering of $Ba_2LuMoO_6$.** Inelastic neutron scattering intensity maps measured with $E_i$ = 70 meV at **a** 7.5 K, **b** 32 K, **c** 75 K and **d** 100 K as function of energy transfer and $|Q|$, corrected for background by subtracting Bose-factor corrected 200K data. A weakly dispersing, gapped singlet-triplet excitation is observed at 28 meV in the 7.5 K and 32 K data. At higher temperatures singlets are no longer observed. **e** Cuts of the $E_i$ = 70 meV Bose-factor corrected inelastic neutron scattering data showing the integrated intensity for 0 < $|Q|$ < 2 Å$^{-1}$ as function of energy transfer. The singlet-triplet excitation centered at 28 meV decreases in intensity with increasing temperature, but no other features are observed. **f** Cut of Bose-factor corrected 7.5 K $E_i$ = 30 meV data minus the Bose-factor corrected 200 K $E_i$ = 30 meV data showing additional excitation features at 11 and 17 meV.

Magnetic excitations in $Ba_2LuMoO_6$ were investigated using inelastic neutron scattering. The nonmagnetic spin singlets expected in a valence bond glass state can be probed using this technique. The inelastic scattering from the sample was weak, and features in the spectra were only visible after subtraction of Bose-factor corrected high-temperature data. The spectra measured at 7.5 K with incident energy $E_i$ = 70 meV are shown in Fig. 5a. The sample spectra are featureless except for a gapped excitation at 28 meV. Note that the feature observed up to 10 meV in this figure is the coherent and incoherent elastic scattering, which is broad in energy transfer on the intensity scale necessary to see this excitation. The excitation at 28 meV gets weaker with increasing $|Q|$ suggesting that it is magnetic in origin. The energy of this excitation is mostly $|Q|$ independent, and it can be interpreted as a weakly dispersing singlet-triplet excitation[28,43]. This indicates that $Ba_2LuMoO_6$ has a singlet ground state with a singlet-triplet gap of $\Delta$ = 28 meV. This is consistent with the singlet-triplet gap measured for $Ba_2YMoO_6$[28].

The singlet-triplet excitation is clearly observed also in the 32 K data (Fig. 5b), showing that the spin singlets persist up to at least this temperature. At 75 K (Fig. 5c) and 100 K (Fig. 5d) the excitation is already difficult to observe. Fig. 5e shows the integrated scattering intensity for $0 < |Q| < 2$ Å$^{-1}$ as a function of energy transfer at different temperatures. The excitation at 28 meV becomes weaker as temperature increases, which is also consistent with it being of magnetic origin. If the feature at 28 meV was related to a crystal field excitation of $Mo^{5+}$, it would not decay this fast with increasing temperature as 28 meV corresponds to over 300 K. Paramagnetic scattering within the gap is not visible in the 70 meV incident energy data, but in the 30 meV data at 7.5 K there is evidence of a far weaker continuum of states peaked at 11 and 17 meV (Fig. 5f), similar to those observed in $Ba_2YMoO_6$[28].

**DISCUSSION**

What is the ground state of $Ba_2LuMoO_6$? The proposed ground states of a $d^1$ double perovskite are different types of magnetic order, spin glass, multipolar order, valence bond glass and spin liquid. A previous neutron diffraction study suggests the material is not magnetically ordered down to 2 K, since no magnetic Bragg peaks were observed[33]. However, as the intensity of the magnetic scattering is proportional to the square of the magnetic moment (< 1 $\mu_B$), establishing the presence or absence of magnetic order using neutron diffraction is challenging in the $d^1$ double perovskites. Our muon spin rotation and relaxation results conclusively rule out magnetic order or spin freezing down to 60 mK. It is unlikely that the ground state involves multipolar order, as the double perovskites with quadrupolar order also develop magnetic order at low temperatures.[14,16,20–23] Moreover, the quadrupolar ordering transition can be observed in the specific heat data as a broad peak, which is not present for $Ba_2LuMoO_6$. Therefore, the main candidate ground states are a valence bond glass state or a spin liquid.

$Ba_2LuMoO_6$ has several properties that support a valence bond glass state. First, the magnetic susceptibility has two Curie-Weiss regions, where the low-temperature region is consistent with the pseudo-gap expected of a valence bond glass state.[26,44] The muon results support a dynamic magnetic ground state, which is consistent with the presence of dynamic orphan spins as expected for a VBG. Finally, we observe a singlet-triplet excitation in inelastic neutron scattering experiments as expected of the dimer singlets of a VBG. Orbital excitations of the valence bond glass state can also explain the weak in-gap scattering in $Ba_2LuMoO_6$[13]. The presence of all these features makes a valence bond glass state the natural explanation for the ground state of $Ba_2LuMoO_6$[13].

The absence of any magnetic order or spin freezing in the muon experiments could also be interpreted as a spin liquid state. Natori *et al.*[9,10] have proposed a chiral spin-orbital liquid as a possible ground state for a $d^1$ double perovskite. The simulated inelastic neutron scattering spectra of such a phase consists of a single broad excitation at a defined energy[10], which is also what we observe for $Ba_2LuMoO_6$. However, it does not explain the weak in-gap inelastic scattering. The proposed gapless spin liquid has a vanishing density of states, which results in a vanishing magnetic susceptibility ($\chi \propto T^{1/2}$) and specific heat ($C_p \propto T^{3/2}$) at low temperatures[9]. This is inconsistent with our results for $Ba_2LuMoO_6$: in fact, we observe a $1/T$ relation for the low-temperature magnetic susceptibility. While we cannot completely rule out the possibility of some type of a spin liquid state in $Ba_2LuMoO_6$, it does not appear to be the proposed chiral spin liquid state.

Finally, disorder in quantum spin systems can induce a random singlet ground state, which has properties similar to those of spin liquids. Random singlet states have mainly been studied on two-dimensional lattices[45–50] and the pyrochlore lattice[51–53], and the authors are unaware of any theoretical predictions of such state on the fcc lattice of $Ba_2LuMoO_6$. A key requirement for a random-singlet state is the presence of disorder. The main type of structural disorder in double perovskites is antisite disorder between the two *B*-sites[8]. Rietveld refinement of our laboratory x-ray diffraction data revealed 1(1)% antisite disorder between $Lu^{3+}$ and $Mo^{5+}$, whereas a previous neutron diffraction study on $Ba_2LuMoO_6$ did not reveal any antisite disorder[33]. This is comparable to $Ba_2YMoO_6$, where 3% antisite disorder was detected by NMR[31]. Whether this would be sufficient disorder to form a random singlet state in the fcc lattice is an open question. The lack of magnetic ordering or spin freezing down to 60 mK is consistent with a random singlet state, as is the upturn in magnetic susceptibility at low temperatures. In terms of inelastic neutron scattering, we do not observe significant scattering at the $|Q|$ positions of the magnetic Bragg peaks of the ordered parent phases as is the case in the $Cu^{2+}$ double perovskite $Sr_2CuTe_{1-x}W_xO_6$[54–56], which has been proposed to have a random singlet ground state[48,49]. Cuts of the elastic line (Supplementary Fig. 2) and various energy ranges (Supplementary Figs. 3, 4, 5) confirm such scattering is not present for $Ba_2LuMoO_6$.

It should be noted that the main theory papers on valence bond glass and spin liquid ground states on fcc lattices assume a $J_{eff} = 3/2$ state on the $Mo^{5+}$, whereas a random singlet state could likely arise in a $S = 1/2$ scenario as well. While the lower than expected paramagnetic moment suggests the presence of a partially unquenched orbital moment as opposed to a pure $S$ = 1/2 state, we are unable to convincingly establish a $J_{eff}$ = 3/2 state based on the experimental data. Given that a theoretical framework for understanding the all physical properties of $Ba_2LuMoO_6$ exists, the valence bond glass state of $J_{eff}$ = 3/2 pseudospins, the authors have chosen to use this interpretation.

The natural comparison for $Ba_2LuMoO_6$ is $Ba_2YMoO_6$, which has also been suggested to be a valence bond glass. Both compounds are cubic $Mo^{5+}$ double perovskites with similar lattice parameters, where a static Jahn-Teller distortion from cubic symmetry is not observed even at 2 K[27,33]. The absence of anomalous oxide displacements, as manifested in anisotropic oxide displacements[33], also rule out any dynamic or disordered local breaking of the $Mo^{5+}$ ligand crystal field in $Ba_2LuMoO_6$. The magnetic susceptibilities of $Ba_2LuMoO_6$ and $Ba_2YMoO_6$ are very similar: two Curie-Weiss regions are observed and interpreted as related to a pseudo-gap of the VBG state. In inelastic neutron scattering, both compounds have a singlet-triplet excitation with the same energy of 28 meV. The main difference is in the muon spin relaxation responses. In $Ba_2LuMoO_6$, we do not observe any static magnetism down to 60 mK. The muon spin relaxation rate of the dynamic low-temperature state is very high, corresponding to exceedingly slow field fluctuations. In comparison, the orphan spins of $Ba_2YMoO_6$ form a dilute spin glass at $T_g$ = 1.3 K, and the muon spin relaxation rate is much lower than in $Ba_2LuMoO_6$. The lack of this spin glass transition makes $Ba_2LuMoO_6$ a more promising system for investigating exotic ground states of the $d^1$ double perovskites.

Recently, the cubic double perovskite $Ba_2Y_xWO_6$ has been proposed to be a $W^{5+}$ $5d^1$ system with a possible valence bond glass state[57]. Muon spin relaxation experiments revealed the lack of magnetic ordering or spin freezing down to 26 mK[57]. A broad maximum was observed in the specific heat similar to $Ba_2YMoO_6$ and $Ba_2LuMoO_6$, and magnetic susceptibility suggested the presence of random magnetism[57]. However, this material is known to have a high number of $Y^{3+}$ vacancies[58,59], which were not considered in the study[57]. The oxidation state of the tungsten cation depends on the yttrium stoichiometry: if $x$ = 2/3, the compound contains only nonmagnetic $W^{6+}$. If $x$ = 1, all tungsten in the material is $W^{5+}$ and it becomes a true $5d^1$ fcc antiferromagnet. Previous neutron diffraction studies put the solubility limit of $Ba_2Y_xWO_6$ at $2/3 \leq x \leq 0.78$[60]. This means that at most 1/3$^{rd}$ of tungsten cations are magnetic $W^{5+}$, and the fcc lattice is significantly diluted. Thus, the lack of magnetic ordering in muon experiments[57] is likely due to the low proportion of magnetic $W^{5+}$ cations in the material. This is supported by the very low effective paramagnetic moment observed at high temperatures[57]. The specific heat data is identical to previous data on $Ba_2Y_{0.78}WO_6$, and the resulting high magnetic entropy can be attributed to a poor lattice match[60]. Due to the uncertainty around stoichiometry, tungsten oxidation state and magnetic dilution in $Ba_2Y_xWO_6$, we will not include it in the following discussion.

We present a comparison of the magnetic properties of related $d^1$ double perovskites in Table I. $Ba_2LuMoO_6$ and $Ba_2YMoO_6$ are the only cubic $Mo^{5+}$ double perovskites with one magnetic cation. $Sr_2YMoO_6$ adopts a monoclinic $P2_1/n$ structure due to the smaller size of the A-site cation[27]. A ferromagnetic transition is observed in the magnetic susceptibility at 8 K, but the saturation magnetization is only 0.12 $\mu_B$ per Mo. Magnetic Bragg peaks were not observed by neutron diffraction, but this could be due to the very weak magnetic scattering from such small moments. $La_2LiMoO_6$ is also monoclinic with a Type I antiferromagnetic ground state and an unusually low ordered moment of 0.32 $\mu_B$[36]. This is likely due to the orbital moment, which in a $d^1$ compound will oppose the spin moment due to spin-orbit coupling. $Sr_2ScMoO_6$ is a tetragonal double perovskite with a high degree of *B*-site disorder between $Mo^{5+}$ and $Sc^{3+}$ [61]. The magnetic susceptibility does not follow the Curie-Weiss law and the magnetic ground state is not known.

The unusual properties of the $d^1$ double perovskites are linked to the spin-orbital entangled $J_{eff}$ = 3/2 pseudospins and their complex interactions. However, structural distortions from cubic symmetry can lift the degeneracy of these states and quench the orbital moment. It is therefore important to compare the cubic $4d^1$ $Ba_2LuMoO_6$ double perovskite to the cubic $5d^1$ analogues, that also retain the $J_{eff}$ = 3/2 pseudospins. Four such cubic compounds are known, where the only magnetic cation is a $5d^1$ cation. $Ba_2MgReO_6$ has canted antiferromagnetic order below $T_N$ = 18 K. More importantly, there is growing evidence of quadrupolar ordering in $Ba_2MgReO_6$ at $T_q$ = 33 K based on specific heat, X-ray and neutron diffraction and resonant inelastic X-ray scattering experiments[21–24]. The quadrupolar ordering is associated with a subtle structural transition from cubic to tetragonal symmetry[23,24]. $Ba_2ZnReO_6$ is similar to $Ba_2MgReO_6$ with a magnetic ordering transition at 16 K and a broad bump in specific heat at 33 K, which could be related to quadrupolar ordering[21,62,63]. However, the type of magnetic ordering is different: $Ba_2ZnReO_6$ is a canted ferromagnet[62], while $Ba_2MgReO_6$ is a canted antiferromagnet[23]. $Ba_2NaOsO_6$ is also a canted ferromagnet but with a lower transition temperature of $T_C$ = 7 K. The magnetic order is thought to be related to orbital ordering due to a quadrupolar transition at $T_q$ = 9.5 K[14–20]. $Ba_2LiOsO_6$ is antiferromagnetic below $T_N$ = 8 K with some disorder in the static fields based on muon experiments[15,64].

The cubic $4d^1$ double perovskites $Ba_2LuMoO_6$ and $Ba_2YMoO_6$ have very similar properties, but they differ greatly from the $5d^1$ analogues that all involve static magnetism, with some compounds also having likely multipolar order. The origin of this difference between $4d^1$ and $5d^1$ double perovskites is not well understood. In general, 5d systems have stronger spin-orbit coupling $\lambda$ and weaker on-site coulombic repulsion $U$ in comparison to 4d systems[1]. The differences in the strength of the spin-orbit coupling $\lambda$ could explain the manifest difference in ground states between the $4d^1$ and $5d^1$ double perovskites. Romhányi, Balents and Jackeli[13] have proposed a microscopic model for the $d^1$ double perovskites. In this model, a weaker $\lambda$ favors a disordered dimer-singlet phase as observed in $4d^1$ Ba$_2$LuMoO$_6$ and Ba$_2$YMoO$_6$, whereas a stronger $\lambda$ favors magnetically ordered phases found in the $5d^1$ analogues Ba$_2$MgReO$_6$ and Ba$_2$NaOsO$_6$. However, questions remain over the role of the Jahn-Teller effects, which are not observed in the $4d^1$ systems, but are thought to drive the quadrupolar ordering in the $5d^1$ systems[2].

Table I. Magnetic properties of double perovskites, where the only magnetic cation is a $d^1$ cation (Mo$^{5+}$ $4d^1$, Re$^{6+}$ $5d^1$ or Os$^{7+}$ $5d^1$) on the B''-site. Adapted and expanded from refs. [12,21]

| Compound | Space group | Ground state | Transition | $\Theta_{CW}$ (K) | $\mu_{eff}$ ($\mu_B$) | Ref. |
|---|---|---|---|---|---|---|
| Ba$_2$YMoO$_6$ | $Fm\bar{3}m$ | VBG + SG | < 50 mK, $T_g$ = 1.3 K | -160 | 1.44 | [25–27] |
| Ba$_2$LuMoO$_6$ | $Fm\bar{3}m$ | VBG | < 60 mK | -114 | 1.32 | This work, [33] |
| Sr$_2$YMoO$_6$ | $P2_1/n$ | FM? | $T_C$ = 8 K | -50 | 1.4 | [27] |
| Sr$_2$ScMoO$_6$ | $I4/m$ | ? | - | - | - | [61] |
| La$_2$LiMoO$_6$ | $P2_1/n$ | Type I AFM | $T_N$ = 18 K | -59 | 1.51 | [31,36] |
| Ba$_2$MgReO$_6$ | $Fm\bar{3}m$ | Canted AFM, quadrupolar | $T_N$ = 18 K, $T_q$ = 33 K | -15 | 0.68 | [21–23] |
| Ba$_2$ZnReO$_6$ | $Fm\bar{3}m$ | Canted FM, quadrupolar? | $T_N$ = 16 K, $T_q$ = 33 K? | -66 | 0.94 | [21,62] |
| Ba$_2$CaReO$_6$ | $I4/m$ | AFM | $T_N$ = 15 K | -39 | 0.74 | [65] |
| Ba$_2$CdReO$_6$ | $I4/m$ | AFM, quadrupolar? | $T_N$ = 12 K, $T_q$ = 25 K? | -15 | 0.72 | [63] |
| Sr$_2$MgReO$_6$ | $I4/m$ | Type I AFM | $T_N$ = 55 K | -134 | 0.8 | [66–68] |
| Sr$_2$CaReO$_6$ | $P2_1/n$ | Spin glass | $T_g$ = 14 K | - | - | [67,69] |
| Ba$_2$LiOsO$_6$ | $Fm\bar{3}m$ | AFM | $T_N$ = 8 K | -40 | 0.73 | [15,64] |
| Ba$_2$NaOsO$_6$ | $Fm\bar{3}m$ | Canted FM, quadrupolar | $T_C$ = 7 K, $T_q$ = 9.5 K | -10 | 0.6 | [14–16,20] |
| Sr$_2$LiOsO$_6$ | $I4/m$ | AFM | $T_N$ = 12 K | -202 | 1.07 | [70] |
| Sr$_2$NaOsO$_6$ | $P2_1/n$ | AFM | $T_N$ = 17 K | -38 | 0.7 | [71] |

In conclusion, our muon and neutron experiments support a valence bond glass ground state in the $4d^1$ fcc antiferromagnet Ba$_2$LuMoO$_6$. Inelastic neutron scattering experiments confirmed the presence of spin singlets with a singlet-triplet gap of Δ = 28 meV. In addition to the nonmagnetic singlets, muon spin rotation and relaxation measurements confirm the presence of dynamic electronic spins down to 60 mK. These results are interpreted as a valence bond glass state, where spin singlet dimers form in a disordered manner and some of the leftover orphan spins remain paramagnetic[13]. However, a quantum spin liquid[10] or a disorder-induced random-singlet state cannot be ruled out. The properties of $Ba_2LuMoO_6$ are very similar to those of isostructural $Ba_2YMoO_6$, which has been proposed to be a valence bond glass. The lack of any spin freezing at low temperatures makes $Ba_2LuMoO_6$ a more promising candidate for such a ground state than $Ba_2YMoO_6$, in which the orphan spins freeze into a spin glass at 1.3 K[29]. The behavior of the cubic $4d^1$ double perovskites $Ba_2LuMoO_6$ and $Ba_2YMoO_6$ is fundamentally different to that of the $5d^1$ analogues $Ba_2MgReO_6$, $Ba_2ZnReO_6$ and $Ba_2NaOsO_6$, which have both magnetic and multipolar order.

**METHODS**

**Sample synthesis and purity**. Polycrystalline powder samples of $Ba_2LuMoO_6$ were prepared by a solid-state reaction method. Stoichiometric amounts of $BaCO_3$ (Alfa Aesar 99.997%), $Lu_2O_3$ (Alfa Aesar 99.99%, pre-treated at 800 °C), and $MoO_3$ (Alfa Aesar 99.998%) were mixed, pelletized and calcined at 800 °C in air for 24 h. The pellets were reground and fired at 1250 °C in flowing 5% $H_2/N_2$ gas (60 cm$^3$/min) for 96 h with intermittent grindings. Phase purity of the samples was investigated using X-ray powder diffraction. The data were collected with a Panalytical X'Pert 3 Powder diffractometer (Cu $K_\alpha$ radiation). Rietveld refinement[72] was carried out using FULLPROF[73] and the crystal structure was visualized using VESTA[74].

**Bulk properties**. Magnetic properties were measured with a Quantum Design MPMS3 SQUID magnetometer. 100 mg of sample powder was enclosed in a gelatin capsule and placed in a plastic straw. The sample shape was estimated as a cylinder of 5 mm diameter and 2 mm length. This corresponds to a moment artefact of 1.072 in a DC measurement on the MPMS3, and all measured moments were divided by this factor. The temperature-dependent magnetization was measured in a field of 1000 Oe from 2 to 300 K in zero-field cool (ZFC) and field cool (FC) modes. Field-dependent magnetization was measured at 2 K between -50 000 and 50 000 Oe. The specific heat of $Ba_2LuMoO_6$ was measured on a Quantum Design Physical Property Measurement System. Sample and silver powders were mixed in a 1:1 ratio and pressed into pellets. Specific heat was then measured on pellet pieces of ≈20 mg mass. The silver contribution was subtracted from the measured specific heat using data from a pressed pellet of pure silver powder.

**Muon spectroscopy**. Muon spin rotation and relaxation (μSR) experiments were carried out at the MUSR instrument at the ISIS Neutron and Muon Source. Approximately 3g of sample powder was attached to a silver sample holder with GE varnish. The sample was cooled in a dilution fridge. Zero-field, transverse field and longitudinal field measurements were carried out between 60 mK and 4 K. The collected data is available online[75].

**Inelastic neutron scattering**. Inelastic neutron scattering was measured on an 8 g $Ba_2LuMoO_6$ sample on the MERLIN spectrometer at ISIS Neutron and Muon Source. The sample powder was contained in a cylindrical aluminium can to give an annular geometry. Measurements were performed with incident energies of 70 and 30 meV at temperatures between 7 and 200 K. The data were reduced using standard Mantid reduction procedures[76]. The inelastic neutron scattering data is available online[77].

**References**

1.  Witczak-Krempa, W., Chen, G., Kim, Y. B. & Balents, L. Correlated Quantum Phenomena in the Strong Spin-Orbit Regime. *Annu. Rev. Condens. Matter Phys.* **5**, 57–82 (2014).
2.  Takayama, T., Chaloupka, J., Smerald, A., Khaliullin, G. & Takagi, H. Spin–Orbit-Entangled Electronic Phases in 4*d* and 5*d* Transition-Metal Compounds. *J. Phys. Soc. Japan* **90**, 062001 (2021).
3.  Takagi, H., Takayama, T., Jackeli, G., Khaliullin, G. & Nagler, S. E. Concept and realization of Kitaev quantum spin liquids. *Nat. Rev. Phys.* **1**, 264–280 (2019).
4.  Browne, A. J., Krajewska, A. & Gibbs, A. S. Quantum materials with strong spin-orbit coupling: challenges and opportunities for materials chemists. *J. Mater. Chem. C* **9**, 11640–11654 (2021).
5.  Kim, B. J. *et al.* Novel $J_{eff}$=1/2 mott state induced by relativistic spin-orbit coupling in $Sr_2IrO_4$. *Phys. Rev. Lett.* **101**, 076402 (2008).
6.  Kitaev, A. Anyons in an exactly solved model and beyond. *Ann. Phys.* **321**, 2–111 (2006).
7.  Khomskii, D. I. & Streltsov, S. V. Orbital Effects in Solids: Basics, Recent Progress, and Opportunities. *Chem. Rev.* **121**, 2992–3030 (2021).
8.  Vasala, S. & Karppinen, M. $A_2B'B''O_6$ perovskites: A review. *Prog. Solid State Chem.* **43**, 1–36 (2015).
9.  Natori, W. M. H., Andrade, E. C., Miranda, E. & Pereira, R. G. Chiral Spin-Orbital Liquids with Nodal Lines. *Phys. Rev. Lett.* **117**, 017204 (2016).
10. Natori, W. M. H., Daghofer, M. & Pereira, R. G. Dynamics of a *j*= 3/2 quantum spin liquid.

Phys. Rev. B **96**, 125109 (2017).

11. Rau, J. G., Lee, E. K. H. & Kee, H. Y. Generic spin model for the honeycomb iridates beyond the Kitaev limit. *Phys. Rev. Lett.* **112**, 077204 (2014).

12. Chen, G., Pereira, R. & Balents, L. Exotic phases induced by strong spin-orbit coupling in ordered double perovskites. *Phys. Rev. B - Condens. Matter Mater. Phys.* **82**, 174440 (2010).

13. Romhányi, J., Balents, L. & Jackeli, G. Spin-Orbit Dimers and Noncollinear Phases in $d^1$ Cubic Double Perovskites. *Phys. Rev. Lett.* **118**, 217202 (2017).

14. Erickson, A. S. *et al.* Ferromagnetism in the Mott Insulator $Ba_2NaOsO_6$. *Phys. Rev. Lett.* **99**, 016404 (2007).

15. Steele, A. J. *et al.* Low-moment magnetism in the double perovskites $Ba_2MOsO_6$ (*M* = Li,Na). *Phys. Rev. B* **84**, 144416 (2011).

16. Lu, L. *et al.* Magnetism and local symmetry breaking in a Mott insulator with strong spin orbit interactions. *Nat. Commun.* **8**, 14407 (2017).

17. Liu, W. *et al.* Phase diagram of $Ba_2NaOsO_6$, a Mott insulator with strong spin orbit interactions. *Phys. B Condens. Matter* **536**, 863–866 (2018).

18. Liu, W., Cong, R., Reyes, A. P., Fisher, I. R. & Mitrović, V. F. Nature of lattice distortions in the cubic double perovskite $Ba_2NaOsO_6$. *Phys. Rev. B* **97**, 224103 (2018).

19. Cong, R., Nanguneri, R., Rubenstein, B. & Mitrović, V. F. Evidence from first-principles calculations for orbital ordering in $Ba_2NaOsO_6$: A Mott insulator with strong spin-orbit coupling. *Phys. Rev. B* **100**, 245141 (2019).

20. Willa, K. *et al.* Phase transition preceding magnetic long-range order in the double perovskite $Ba_2NaOsO_6$. *Phys. Rev. B* **100**, 041108 (2019).

21. Marjerrison, C. A. *et al.* Cubic $Re^{6+}$ ($5d^1$) Double Perovskites, $Ba_2MgReO_6$, $Ba_2ZnReO_6$, and $Ba_2Y_{2/3}ReO_6$ : Magnetism, Heat Capacity, μSR, and Neutron Scattering Studies and Comparison with Theory. *Inorg. Chem.* **55**, 10701–10713 (2016).

22. Hirai, D. & Hiroi, Z. Successive Symmetry Breaking in a $J_{eff}$ = 3/2 Quartet in the Spin–Orbit Coupled Insulator $Ba_2MgReO_6$. *J. Phys. Soc. Japan* **88**, 064712 (2019).

23. Hirai, D. *et al.* Detection of multipolar orders in the spin-orbit-coupled $5d$ Mott insulator $Ba_2MgReO_6$. *Phys. Rev. Res.* **2**, 022063 (2020).

24. Lovesey, S. W. & Khalyavin, D. D. Magnetic order and $5d^1$ multipoles in a rhenate double perovskite $Ba_2MgReO_6$. *Phys. Rev. B* **103**, 235160 (2021).

25. Cussen, E. J., Lynham, D. R. & Rogers, J. Magnetic Order Arising from Structural Distortion: Structure and Magnetic Properties of $Ba_2LnMoO_6$. *Chem. Mater.* **18**, 2855–2866 (2006).

26. De Vries, M. A., Mclaughlin, A. C. & Bos, J.-W. G. Valence Bond Glass on an fcc Lattice in the

Double Perovskite $Ba_2YMoO_6$. *Phys. Rev. Lett.* **104**, 177202 (2010).

27. Mclaughlin, A. C., De Vries, M. A. & Bos, J. W. G. Persistence of the valence bond glass state in the double perovskites $Ba_{2-x}Sr_xYMoO_6$. *Phys. Rev. B - Condens. Matter Mater. Phys.* **82**, 094424 (2010).

28. Carlo, J. P. *et al.* Triplet and in-gap magnetic states in the ground state of the quantum frustrated fcc antiferromagnet $Ba_2YMoO_6$. *Phys. Rev. B - Condens. Matter Mater. Phys.* **84**, 100404 (2011).

29. De Vries, M. A. *et al.* Low-temperature spin dynamics of a valence bond glass in $Ba_2YMoO_6$. *New J. Phys.* **15**, 043024 (2013).

30. Qu, Z. *et al.* Spin-phonon coupling probed by infrared transmission spectroscopy in the double perovskite $Ba_2YMoO_6$. *J. Appl. Phys.* **113**, 2011–2014 (2013).

31. Aharen, T. *et al.* Magnetic properties of the geometrically frustrated $S=½$ antiferromagnets, $La_2LiMoO_6$ and $Ba_2YMoO_6$, with the B-site ordered double perovskite structure: Evidence for a collective spin-singlet ground state. *Phys. Rev. B* **81**, 224409 (2010).

32. Streltsov, S. V. & Khomskii, D. I. Jahn-Teller Effect and Spin-Orbit Coupling: Friends or Foes? *Phys. Rev. X* **10**, 031043 (2020).

33. Coomer, F. C. & Cussen, E. J. Structural and magnetic properties of $Ba_2LuMoO_6$: A valence bond glass. *J. Phys. Condens. Matter* **25**, 082202 (2013).

34. Koseki, S., Matsunaga, N., Asada, T., Schmidt, M. W. & Gordon, M. S. Spin-Orbit Coupling Constants in Atoms and Ions of Transition Elements: Comparison of Effective Core Potentials, Model Core Potentials, and All-Electron Methods. *J. Phys. Chem. A* **123**, 2325–2339 (2019).

35. Iwahara, N., Vieru, V. & Chibotaru, L. F. Spin-orbital-lattice entangled states in cubic d1 double perovskites. *Phys. Rev. B* **98**, 075138 (2018).

36. Dragomir, M. *et al.* Magnetic ground state of $La_2LiMoO_6$: A comparison with other $Mo^{5+}$ ($S = 1/2$) double perovskites. *Phys. Rev. Mater.* **4**, 104406 (2020).

37. Yamashita, S., Nakazawa, Y., Ueda, A. & Mori, H. Thermodynamics of the quantum spin liquid state of the single-component dimer Mott system κ-$H_3$(Cat-EDT-TTF)$_2$. *Phys. Rev. B* **95**, 184425 (2017).

38. de Réotier, P. D. & Yaouanc, A. Muon spin rotation and relaxation in magnetic materials. *J. Phys. Condens. Matter* **9**, 9113–9166 (1997).

39. Mendels, P. *et al.* Quantum Magnetism in the Paratacamite Family: Towards an Ideal Kagomé Lattice. *Phys. Rev. Lett.* **98**, 077204 (2007).

40. Tustain, K. *et al.* From magnetic order to quantum disorder in the Zn-barlowite series of $S = 1/2$ kagomé antiferromagnets. *npj Quantum Mater.* **5**, 74 (2020).

41. Mustonen, O. *et al.* Spin-liquid-like state in a spin-1/2 square-lattice antiferromagnet perovskite induced by $d^{10}$–$d^0$ cation mixing. *Nat. Commun.* **9**, 1085 (2018).
42. Mustonen, O. *et al.* Tuning the $S = 1/2$ square-lattice antiferromagnet $Sr_2Cu(Te_{1-x}W_x)O_6$ from Néel order to quantum disorder to columnar order. *Phys. Rev. B* **98**, 064411 (2018).
43. Kageyama, H. *et al.* Direct evidence for the localized single-triplet excitations and the dispersive multitriplet excitations in $SrCu_2(BO_3)_2$. *Phys. Rev. Lett.* **84**, 5876–5879 (2000).
44. Tarzia, M. & Biroli, G. The valence bond glass phase. *EPL (Europhysics Lett.* **82**, 67008 (2008).
45. Watanabe, K., Kawamura, H., Nakano, H. & Sakai, T. Quantum Spin-Liquid Behavior in the Spin-1/2 Random Heisenberg Antiferromagnet on the Triangular Lattice. *J. Phys. Soc. Japan* **83**, 034714 (2014).
46. Shimokawa, T., Watanabe, K. & Kawamura, H. Static and dynamical spin correlations of the $S$=1/2 random-bond antiferromagnetic Heisenberg model on the triangular and kagome lattices. *Phys. Rev. B - Condens. Matter Mater. Phys.* **92**, 134407 (2015).
47. Uematsu, K. & Kawamura, H. Randomness-Induced Quantum Spin Liquid Behavior in the $s = 1/2$ Random $J_1$–$J_2$ Heisenberg Antiferromagnet on the Honeycomb Lattice. *J. Phys. Soc. Japan* **86**, 044704 (2017).
48. Uematsu, K. & Kawamura, H. Randomness-induced quantum spin liquid behavior in the $s = 1/2$ random $J_1$-$J_2$ Heisenberg antiferromagnet on the square lattice. *Phys. Rev. B* **98**, 134427 (2018).
49. Liu, L., Shao, H., Lin, Y.-C., Guo, W. & Sandvik, A. W. Random-Singlet Phase in Disordered Two-Dimensional Quantum Magnets. *Phys. Rev. X* **8**, 041040 (2018).
50. Kawamura, H. & Uematsu, K. Nature of the randomness-induced quantum spin liquids in two dimensions. *J. Phys. Condens. Matter* **31**, 504003 (2019).
51. Savary, L. & Balents, L. Disorder-Induced Quantum Spin Liquid in Spin Ice Pyrochlores. *Phys. Rev. Lett.* **118**, 087203 (2017).
52. Uematsu, K. & Kawamura, H. Randomness-Induced Quantum Spin Liquid Behavior in the $s$=1/2 Random-Bond Heisenberg Antiferromagnet on the Pyrochlore Lattice. *Phys. Rev. Lett.* **123**, 087201 (2019).
53. Bowman, D. F. *et al.* Role of defects in determining the magnetic ground state of ytterbium titanate. *Nat. Commun.* **10**, 637 (2019).
54. Katukuri, V. M. *et al.* Exchange Interactions Mediated by Non-Magnetic Cations in Double Perovskites. *Phys. Rev. Lett.* **124**, 077202 (2020).
55. Hu, X. *et al.* Freezing of a Disorder Induced Spin Liquid with Strong Quantum Fluctuations. *Phys. Rev. Lett.* **127**, 017201 (2021).

56. Fogh, E. *et al.* Randomness and Frustration in a $S = 1/2$ Square-Lattice Heisenberg Antiferromagnet. *arxiv*:2112.03312 (2021).

57. Lee, S. *et al.* Experimental evidence for a valence-bond glass in the $5d^1$ double perovskite $Ba_2YWO_6$. *Phys. Rev. B* **103**, 224430 (2021).

58. Rauser, G. & Kemmler-Sack, S. Strukturbestimmungen an geordneten Perowskiten des Typs $Ba_2BM^{VI}O_6$. *Zeitschrift für Anorg. und Allg. Chemie* **429**, 181–184 (1977).

59. Schittenhelm, H.-J. & Kemmler-Sack, S. Über geordnete Perowskite mit Kationenfehlstellen Die Systeme $Ba_2MgWO_6$-$Ba_2Y_{0,67}WO_6$ und $Ba_2CaWO_6$-$Ba_2Y_{0,67}WO_6$. *Zeitschrift für Anorg. und Allg. Chemie* **431**, 144–152 (1977).

60. Burrows, O. Structural and Mangetic Properties of the Geometrically Frustrated 3*d* and 5*d* $S$=1/2 Double Perovskites $Sr_2CuWO_6$, $Ba_2YWO_6$ and $LaSrMgWO_6$. (University of Edinburgh, 2016).

61. Wallace, T. K. & McLaughlin, A. C. Structural and magnetic characterisation of the novel spin frustrated double perovskite $Sr_2ScMoO_6$. *J. Solid State Chem.* **219**, 148–151 (2014).

62. da Cruz Pinha Barbosa, V. *et al.* The Impact of Structural Distortions on the Magnetism of Double Perovskites Containing $5d^1$ Transition-Metal Ions. *Chem. Mater.* https://doi.org/10.1021/acs.chemmater.1c03456 (2022)

63. Hirai, D. & Hiroi, Z. Possible quadrupole order in tetragonal $Ba_2CdReO_6$ and chemical trend in the ground states of $5d^1$ double perovskites. *J. Phys. Condens. Matter* **33**, 135603 (2021).

64. Stitzer, K. E., Smith, M. D. & Zur Loye, H. C. Crystal growth of $Ba_2MOSO_6$ (*M* = Li, Na) from reactive hydroxide fluxes. *Solid State Sci.* **4**, 311–316 (2002).

65. Yamamura, K., Wakeshima, M. & Hinatsu, Y. Structural phase transition and magnetic properties of double perovskites $Ba_2CaMO_6$ (*M*=W, Re, Os). *J. Solid State Chem.* **179**, 605–612 (2006).

66. Wiebe, R. *et al.* Frustration-driven spin freezing in the $S$=1/2 fcc perovskite $Sr_2MgReO_6$. *Phys. Rev. B* **68**, 134410 (2003).

67. Greedan, J. E., Derakhshan, S., Ramezanipour, F., Siewenie, J. & Proffen, T. H. A search for disorder in the spin glass double perovskites $Sr_2CaReO_6$ and $Sr_2MgReO_6$ using neutron diffraction and neutron pair distribution function analysis. *J. Phys. Condens. Matter* **23**, 164213 (2011).

68. Gao, S. *et al.* Antiferromagnetic long-range order in the $5d^1$ double-perovskite $Sr_2MgReO_6$. *Phys. Rev. B* **101**, 220412 (2020).

69. Wiebe, C. R., Greedan, J. E., Luke, G. M. & Gardner, J. S. Spin-glass behavior in the $S = 1/2$ fcc ordered perovskite $Sr_2CaReO_6$. *Phys. Rev. B* **65**, 144413 (2002).

70. Feng, H. L. *et al.* Crystal Structure and Magnetic Properties of $Sr_2LiOsO_6$. *JPS Conf. Proc.* **1**, 012002 (2014).

71. Thakur, G. S., Felser, C. & Jansen, M. Structure and Magnetic Properties of $Sr_2NaOsO_6$. *Eur. J. Inorg. Chem.* **2020**, 3991–3995 (2020).

72. Rietveld, H. M. A profile refinement method for nuclear and magnetic structures. *J. Appl. Crystallogr.* **2**, 65–71 (1969).

73. Rodríguez-Carvajal, J. Recent advances in magnetic structure determination by neutron powder diffraction. *Physica B* vol. 192 55–69 (1993).

74. Momma, K. & Izumi, F. VESTA 3 for three-dimensional visualization of crystal, volumetric and morphology data. *J. Appl. Crystallogr.* **44**, 1272–1276 (2011).

75. Cussen, E., Mutch, H., Mustonen, O., Baker, P. & Pughe, C. Investigating the ground state of the proposed valence bond glass $Ba_2LuMoO_6$, STFC ISIS Neutron and Muon Source, https://doi.org/10.5286/ISIS.E.RB1910349 (2019).

76. Arnold, O. *et al.* Mantid - Data analysis and visualization package for neutron scattering and μSR experiments. *Nucl. Instruments Methods Phys. Res. Sect. A Accel. Spectrometers, Detect. Assoc. Equip.* **764**, 156–166 (2014).

77. Walker, H., Cussen, E. & Coomer, F. Inelastic neutron scattering study of a quantum frustrated undistorted fcc lattice with strong spin-orbit coupling, STFC ISIS Neutron and Muon Source, https://doi.org/10.5286/ISIS.E.RB1510299 (2015).

**Data availability**

All data supporting the conclusions of this article are available from the authors upon reasonable request. The collected muon data is available at ref. [75]. The inelastic neutron scattering data is available at ref. [77].

**Acknowledgements**

O.M., H.M., C.P. and E.J.C. are grateful for support from the Leverhulme Trust Research Project Grant RPG-2017-109. S.E.D. acknowledges funding from the Winton Programme for the Physics of Sustainability (Cambridge) and EPSRC (EP/T028580/1). The authors thank the Science and Technology Facilities Council for the beamtime allocated at ISIS. The authors are grateful for access to the MPMS3 instrument at the ISIS Materials Characterisation Laboratory. Heat Capacity measurements were performed using the Advanced Materials Characterisation Suite, funded by EPSRC Strategic Equipment Grant EP/M000524/1. The authors thank Dr Daigorou Hirai for fruitful discussions on fitting the lattice specific heat.

**Author contributions**

O.M., H.M., F.C.C. and E.J.C. planned and conceived the study. E.J.C. supervised the project. H.M. synthesized the samples. O.M., H.M. and G.B.G.S. measured the magnetic susceptibility. C.L. and S.E.D. measured the specific heat. O.M., H.M., P.J.B. and C.P. measured the muon spin rotation and relaxation with analysis by O.M. and H.M. H.W. and R.S.P. measured inelastic neutron scattering. O.M. wrote the manuscript with contributions from all authors.

**Competing interests**

The authors declare no competing interests.

**Materials and correspondence**

Correspondence and requests for materials should be addressed to E.J.C. at e.j.cussen@sheffield.ac.uk

# Supplementary Information

# Valence bond glass state in the 4d$^1$ fcc antiferromagnet Ba$_2$LuMoO$_6$

O. Mustonen,[1] H. Mutch,[1] H. C. Walker,[2] P. J. Baker,[2] F. C. Coomer,[3] R. S. Perry,[4] C. Pughe,[1] G. B. G. Stenning,[2] C. Liu,[5] S. E. Dutton,[5] E. J. Cussen[1*]

## X-ray Diffraction

Laboratory X-ray diffraction was used to evaluate the phase purity and crystal structure of our Ba$_2$LuMoO$_6$ sample. The material crystallises in the cubic $Fm\bar{3}m$ space group corresponding to an ordered double perovskite structure. No additional impurity peaks are observed in the diffraction pattern (Fig. S1).

Results of our Rietveld refinement are presented in Table S1. The refined structure is in agreement with reported low-temperature neutron diffraction data[1]. The degree of *B*-site cation order was evaluated by refining Mo and Lu site occupancies with the total occupancies of Mo and Lu constrained to nominal stoichiometry and Mo on the Lu-site constrained to be the same as Lu on the Mo-site. Atomic displacement parameters of both *B*-sites were constrained to be the same. The displacement parameters are similar to those of Ba$_2$YMoO$_6$ at room temperature[2].

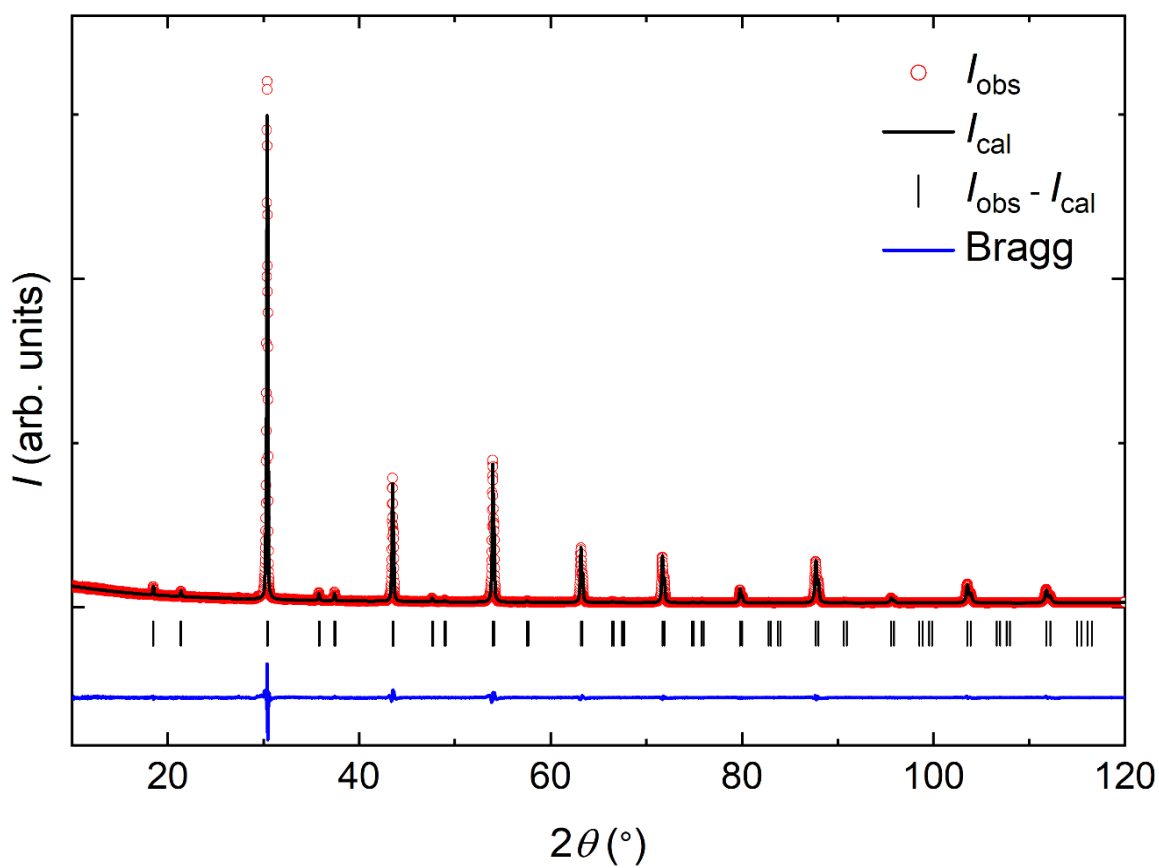

Fig. S1. Laboratory X-ray powder diffraction pattern of Ba$_2$LuMoO$_6$ at room temperature. No impurity peaks are observed.

Table S1. Refined crystal structure of $Ba_2LuMoO_6$ at room temperature. Space group $Fm\bar{3}m$ with $a$ = 8.32058(3) Å, $R_p$ = 13.7%, $R_{wp}$ = 10.8% and $\chi^2$ = 3.05.

| Atom | x | y | z | $B_{iso}$ (Å²) | Occupancy |
|---|---|---|---|---|---|
| Ba | 0.25 | 0.25 | 0.25 | 0.50(4) | 1.00 |
| Lu1 | 0 | 0 | 0 | 0.41(4) | 0.99(1) |
| Mo1 | 0 | 0 | 0 | 0.41(4) | 0.01(1) |
| Mo2 | 0.5 | 0.5 | 0.5 | 0.41(4) | 0.99(1) |
| Lu2 | 0.5 | 0.5 | 0.5 | 0.41(4) | 0.01(1) |
| O | 0.2633(7) | 0 | 0 | 0.7(1) | 1.00 |

**Inelastic neutron scattering**

Our inelastic neutron scattering data shows a clear flat excitation at 28 meV. We have interpreted this as a singlet-triplet excitation consistent with a valence bond glass ground state. Another possible ground state for $Ba_2LuMoO_6$ is the disorder-induced random singlet state. The most structurally similar random singlet candidate is $Sr_2CuTe_{1-x}W_xO_6$, which is also a double perovskite. In $Sr_2CuTe_{1-x}W_xO_6$, significant scattering is observed at the $|Q|$ positions of the Bragg peaks of the magnetically ordered parent phases.[3] This scattering is very noticeable at $|Q|$ ~0.7-0.9 at low energies.

We do not observe such magnetic scattering for $Ba_2LuMoO_6$. The $|Q|$ positions of an ordered fcc antiferromagnet with the lattice parameter of $Ba_2LuMoO_6$ correspond to $|Q|$ = 0.75 Å⁻¹ for Type I order and $|Q|$ = 0.65 Å⁻¹ for Type II order. We do not observe any magnetic scattering at these positions in cuts of the elastic line (Fig. S2), in cuts between 3 meV < E < 5 meV (Fig. S3), cuts between 5 meV < E < 8 meV (Fig. S3) or in cuts between 8 meV < E < 12 meV.

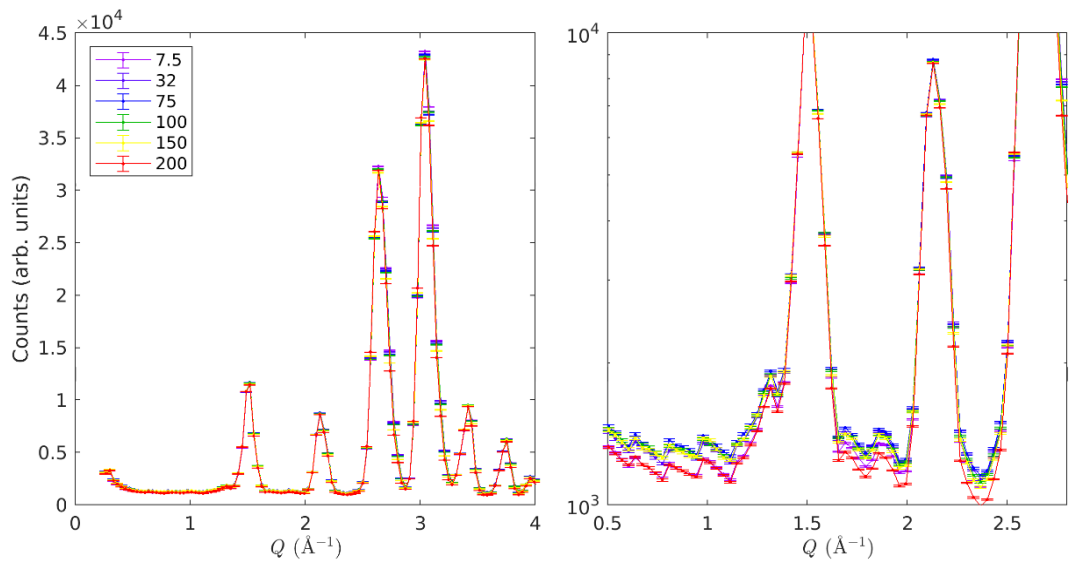

Fig. S2. Cuts of the elastic line from $E_i$ = 30 meV data MERLIN data at different temperatures. No magnetic Bragg peaks are observed. The nuclear scattering from the elastic line bleeds into the inelastic data at low energies, which is why we present it for comparison.

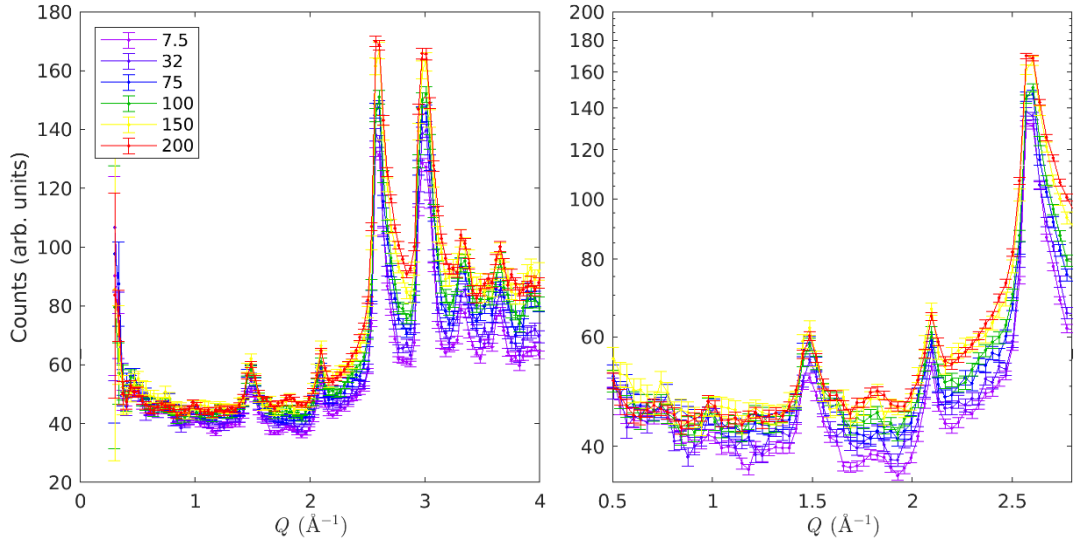

Fig. S3. Cuts of the $E_i$ = 30 meV MERLIN data between 3 and 5 meV at different temperatures. We do not observe magnetic scattering at the positions expected from a magnetically ordered fcc antiferromagnet at $|Q|$ = 0.75 Å$^{-1}$ or $|Q|$ = 0.65 Å$^{-1}$.

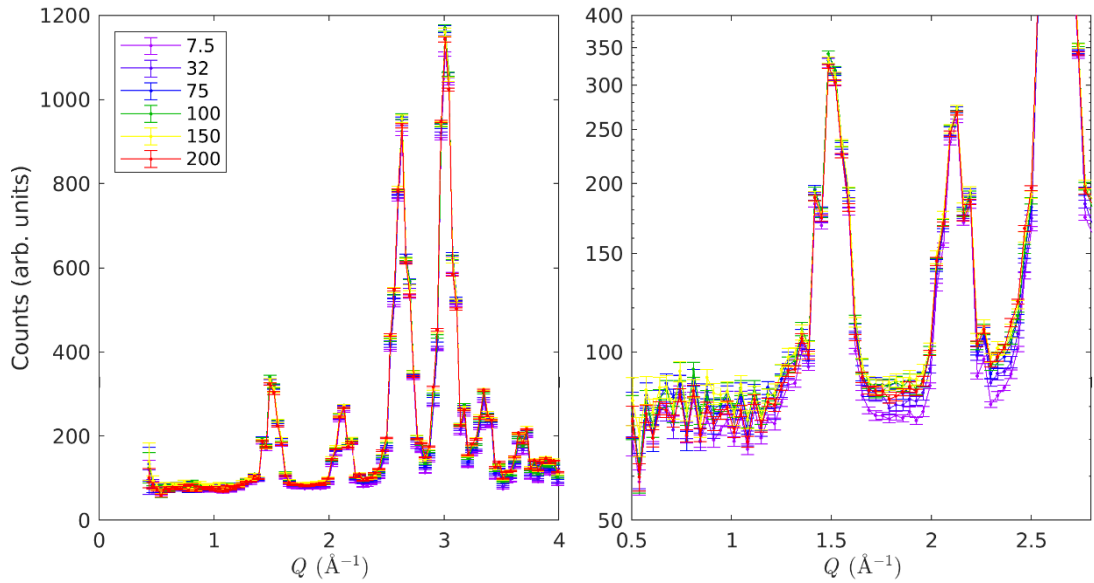

Fig. S4. Cuts of the $E_i$ = 70 meV MERLIN data between 5 and 8 meV at different temperatures. We do not observe magnetic scattering at the positions expected from a magnetically ordered fcc antiferromagnet at $|Q|$ = 0.75 Å$^{-1}$ or $|Q|$ = 0.65 Å$^{-1}$.

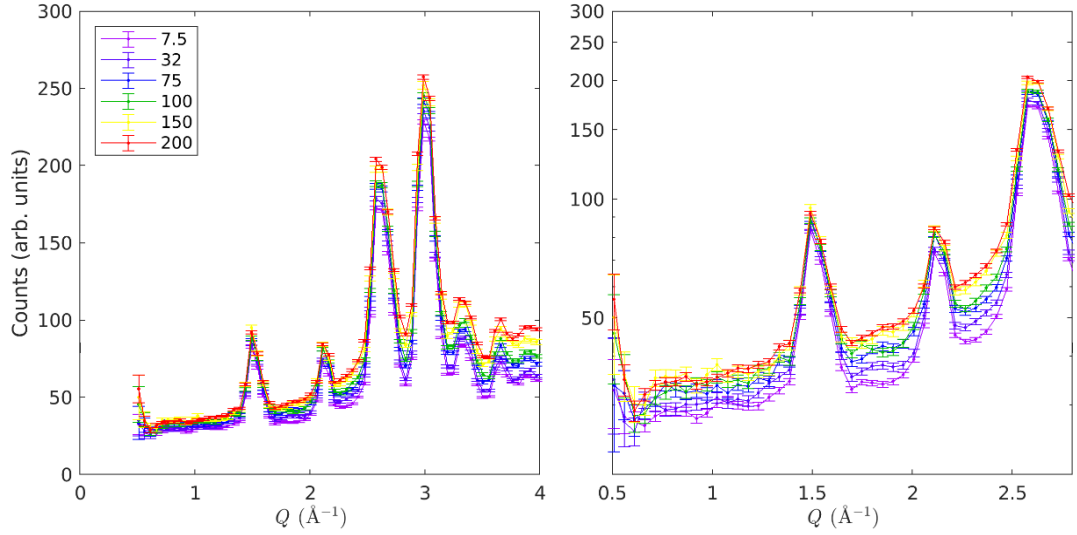

Fig. S5. Cuts of the $E_i$ = 70 meV MERLIN data between 8 and 12 meV at different temperatures. We do not observe magnetic scattering at the positions expected from a magnetically ordered fcc antiferromagnet at $|Q|$ = 0.75 Å$^{-1}$ or $|Q|$ = 0.65 Å$^{-1}$.

**References**

1. Coomer, F. C. & Cussen, E. J. Structural and magnetic properties of $Ba_2LuMoO_6$: A valence bond glass. *J. Phys. Condens. Matter* **25**, 082202 (2013).

2. Mclaughlin, A. C., De Vries, M. A. & Bos, J. W. G. Persistence of the valence bond glass state in the double perovskites $Ba_{2-x}Sr_xYMoO_6$. *Phys. Rev. B - Condens. Matter Mater. Phys.* **82**, 094424 (2010).

3. Fogh, E. *et al.* Randomness and Frustration in a *S* = 1/2 Square-Lattice Heisenberg Antiferromagnet. *arxiv* 2112.03312 (2021).